# Effect of Dielectric Properties of Ceramic-Solvent Interface on the Binding of Protein to Oxide Ceramics: a Non Local Electrostatic Approach

A.I. Rubinstein[1*], R. F. Sabirianov[2*], F. Namavar[1]

[1]Department of Orthopaedic Surgery and Rehabilitation, University of Nebraska Medical Center, Omaha, NE 68198; [2]Department of Physics, University of Nebraska at Omaha, Omaha, NE 68182

**Abstract**
The contribution of electrostatic interactions to the free energy of binding between model protein and a ceramic implant surface in the aqueous solvent, considered in the framework of the non-local electrostatic model, is calculated as a function of the implant low-frequency dielectric constant. We show that the existence of a dynamically ordered (low-dielectric) interfacial solvent layer at the protein-solvent and ceramic-solvent interface markedly increases charging energy of the protein and ceramic implant, and consequently makes the electrostatic contribution to the protein-ceramic binding energy more favorable (attractive). Our analysis shows that the corresponding electrostatic energy between protein and oxide ceramics depends non-monotonically on the dielectric constant of ceramic, $\varepsilon_C$. Obtained results indicate that protein can attract electrostatically to the surface if ceramic material has a moderate $\varepsilon_C$ below or about 35 (in particularly $ZrO_2$ or $Ta_2O_5$). This is in contrast to classical (local) consideration of the solvent, which demonstrates an unfavorable electrostatic interaction of protein with typical metal oxide ceramic materials ($\varepsilon_C > 10$). Thus, a solid implant coated by combining oxide ceramic with a reduced dielectric constant can be beneficial to strengthen the electrostatic binding of the protein-implant complex.

**Introduction**

The response of biological cells to artificial implant surfaces critically depends on their morphology (at the nano–scale), charge distribution, and physicochemical properties that strongly affect the molecular and macromolecular interactions that take place at the implant interface[1,2,3]. Because cells can adhere and function on the implant surface with the help of the extracellular matrix (ECM) proteins,[4,5] understanding mechanisms of the adhesive protein adsorption or formation of the protein-implant surface complexes on the nano-scale level is the key objective in the interdisciplinary study of the biocompatibility and bio-integration of implanted devices. Despite the importance of the above problem, the analysis of the protein adsorption on inorganic surfaces remains a challenge[6,7,8]. One of the very important components of the adsorption energy is attractive electrostatic interactions (EI) between a protein and the surface. Nevertheless its estimation is rather poorly studied due to the absence an adequate consideration of the dielectric response at the protein-solvent and implant-solvent interfaces that determine EI in the protein-solvent-implant interfacial system at the distances comparable with the dynamic solvent microstructure[6,9]. Thus, an accurate description of EI between the protein and solid implant in the aqueous solvent during the adsorption process is needed.

When a dielectric implant surface has effective distributed charges the protein-implant association should be governed mainly by EI prior to establishing substantial surface contacts. There is a competition between contribution of the long-range favorable (attractive) EI and the unfavorable contribution due to electrostatic desolvation of both protein and implant to the association binding energy. The latter contribution is a result of an exclusion of the high-dielectric aqueous solvent when a low-dielectric protein approaches the implant surface (glasses, polymers, ceramics, and other). Thus, the displacement of water molecules from implant surface presents a substantial energy barrier to protein adsorption[10,11]. As a consequence, contribution of EI to the binding energy should be significantly large to compensate or exceed the desolvation effect to facilitate the formation of the protein-implant complex. At the same time, in traditional classic (CL) consideration of the aqueous solvent, the electrostatic contribution to the protein adsorption energy is not significant even so due to the strong electrostatic screening by the aqueous solvent with a high dielectric constant (~80). Thus, a substantial electrostatic binding protein and solid implant is far from evidence. However, several studies[8,12,13,14] (see also Ref.1 and references cited therein) show that protein adsorption does occur on surfaces of hydrophilic materials, where EI provide the necessary driving force for protein adsorption.

Variety of implant materials are commonly used in clinical medicine. Among them gold, carbon and related materials are conducting, while many of biocompatible implants made of dielectric materials such as polyethylene and oxide ceramics including zirconia, titania and tantalum oxide due to their excellent biocompatibility. Typically, the surface of widely used metallic implants such as titanium is oxidized. It has been shown that a titanium surface oxidizes (forms predominantly as dioxide in rutile structure) has the depth of oxidation up to about 8 nm[15,16]. Modifying the surface oxide layers by coating them with ceramics such as zirconia and titania may further improve the properties of implant materials. Particularly, nanostructuring the coating of orthopedic implants can drastically affect their mechanical, dielectric, and wetting properties[17]. Oxide ceramics may be prepared to have similar surface topology and, therefore, this allows comparative experimental studies of protein-surface interaction by addressing the effect of selected property, e.g. dielectric constant, varying in these materials.

The implant dielectric constant, $\varepsilon_C$, of the above dielectrics can vary in range from about 2 in polyethylene to above 300 for $SrTiO_3$. This constant can be selectively tuned in some oxide ceramics to modify the dielectric properties of the surface oxide layer (e.g. $TiO_2$). Titania-based ceramics may have a dielectric constant anywhere between 25 and 114[18]. It is known that surface defects of solid state insulators such as vacancies, steps and edges may have charged states[19]. The interactions between these effective charges and the charged patches of protein depend on the dielectric nature of the implants. It is important to reveal such solid implants that provide the best conditions for forming electrostatically induced protein-implant association.

Experimentally, the amount of protein adsorbed on the surface can be evaluated by ELISA (enzyme-linked immunosorbent assay)[24], fluorescent labeling in PSIM (protein-surface interaction microarrays)[20], radiolabeling or related methods. These methods provide quantitative measure of protein binding affinity to the surface as function of solvent pH or surface parameters (such as roughness). However, these methods do not provide direct information on the binding strength of macromolecules to the surface. In combination with Raman spectroscopy (or Surface Enhanced Raman Spectroscopy) a fingerprints of specific bonding can be identified. Force spectroscopy (based on AFM) was used to measure, in aqueous environment, the dispersion force due to the polymer adsorbed on the substrate. The latter was applied to analysis of protein adsorption on charged surfaces. For example, Meadows et al. (and referenced therein) show that the loading-rate plots contains only a single rupture, which most likely probe a specific surface-protein (fibronectin) interaction[21]. This allows finding the value of the free energy for macromolecule dissociation from the surface. This method allows, in principle, measuring the trends of macromolecule-surface interaction depending on surface parameters (such as charge, preparation conditions, or dielectric constant).

Both experimental studies[22,23] and results of simulations[24,25] show that not only immobilization but also activation (partial unfolding) of the adhesive proteins (such as fibronectin) is affected by negative charges on silica, mica, and other hydrophilic (oxide) surfaces as well as onto the nano-structured surface. These electrostatically induced events should promote ECM formation[4,5,26] that governs cell adhesion and, as a result, cell functioning on the implant surface. Understanding protein activation associated with the general problem of protein partial unfolding, particularly at the inorganic surfaces, is one of the fundamental goals in modern computational bio-chemistry, biomedical sciences, and nanotechnology[1,27,28]. Although there is great research interest to the electrostatic component of the protein adsorption and activation at the inorganic surface, relevant EI estimations are still poor and frequently restricted to CL electrostatics models[1,3,29].

The electrostatic free energy ($\Delta G_{EI}$) of binding between protein and implant can be calculated as corresponding energy for two interacting macromolecules, in particular protein P and ceramic particle C associated into the complex CP. This energy in an aqueous solvent can be computed (analogously to a protein-protein complex) as the difference in the electrostatic free energy between bound ($\Delta G^{CP}$) and unbound protein ($\Delta G^{P}$) and ceramic particle ($\Delta G^{C}$) states[30]:

$$\Delta G_{EI} = \Delta G^{CP} - (\Delta G^{C} + \Delta G^{P}) = \frac{1}{2}\left[\sum_{i \in (C,P)} (q_i \Phi_i)^{CP} - \sum_{i \in C} (q_i \Phi_i)^{C} - \sum_{i \in P} (q_i \Phi_i)^{P}\right], \qquad (1)$$

where the summation is over all the charges $q_i$ located in the macromolecule(s), and $\Phi_i$ is the electrostatic potential at the position of $q_i$. The electrostatic potential can be calculated using the Poisson-Boltzmann equation in the framework of CL electrostatics with a local relation between the field and its electric induction [30,31,32]. Commonly, this approach is treating the interfacial solvent water as bulk, i.e. assuming that its dielectric constant to be ~ 80. However, there is growing number of experimental[33,34] and theoretical[35,36] studies pointing to the existence of a partially structured interfacial water layer with dielectric properties distinct from the bulk solvent. Due to the low rotational mobility of water dipole molecules of the solvent at a dielectric-solvent interface (where dielectric is a protein-like or ceramic medium), the effective dielectric permittivity of the layer is significantly reduced[37,38,39]. Thus, it is necessary to develop more adequate approaches for EI evaluations that can accurately take the above spatial heterogeneity of the solvent into account. Recently, a continuum non-local (NL) electrostatic model, adopting an integral relationship between the electric field and its induction, was developed to analyze the effects of the spatial heterogeneity of the effective dielectric properties at the protein-solvent interface [9,32,40]. Particularly, it was shown that the presence of an extended dynamically ordered water shell at the dielectric-solvent interface with substantially reduced effective dielectric permittivity resulted in the prominent decline of the electric field screening in the vicinity of the interface. This model explains the high magnitude of the protein association rate constants in the protein kinetics [40]. Recently, we have shown that the electrostatic contribution of the protein binding free energy is much stronger (in comparison with CL consideration) due to the existence of the low-dielectric interfacial water shell on the surface of the protein-like dielectric media[9]. This effect can be a major factor capable of compensation for the desolvation effect in the formation of the protein complex. A similar statement can be made about the protein-implant binding free energy during the initial protein adsorption on the inorganic (ceramic implant) surfaces.

In our study, to assess the role of the electrostatic contribution to the binding free energy of protein-ceramic association, $\Delta G_{EI}$, we considered a simple system that is composed of two model macromolecules (represented by continuous medium), in particular, protein P and ceramic C. Single point charges of opposite signs are located at their respective surfaces in the region of functional patches (binding sites). The bound state of the protein-ceramic complex, CP, assumes the contact of these particles at the location of opposite charges by their binding sites. To estimate the electrostatic free energy of the point charge placed into the isolated particle (P and C) in close proximity to the solvent interface ($\Delta G^P$, $\Delta G^C$), we calculated the corresponding "charge image potential" (IP) acting on the above charge due to its electrostatic interaction with the induced polarization at the interface[9]. We used the concept of the NL electrostatics and phenomenological theory of polar solvent[32,40] (and references cited therein). Assuming that the charges located at the particle binding sites are in close proximity to their molecular surface, the interface in the vicinity of the charge was approximated as a locally flat (planar) solvated region.

Here, we analyze the electrostatic free energy $\Delta G_{EI}$ of the model protein-ceramic binding for NL and CL solvent consideration as a function of the ceramic dielectric constant $\varepsilon_C$. Our calculations shows that taking into account the NL electrostatic effect of the solvent (the low-dielectric interfacial water shell on the dielectric interface) significantly shifts (to the negative values) the electrostatic contribution to the binding free energy of the protein-ceramic association, $\Delta G_{EI}$, compared to the CL solvent model. This results in the strong favorable electrostatic contribution to the protein-ceramic complex formation for a wide range of ceramic materials. In particular, $\Delta G_{EI}$ calculated for the protein complex on the $ZrO_2$ surface is more favorable than for other ceramic surfaces including known biocompatible materials such as $Ta_2O_5$ and $TiO_2$.

**METHODS**

We analyzed the electrostatic interactions at the model dielectric-solvent interface using NL electrostatic approach[9, 32., 40]. In this approach, the linear dielectric response for each of the media in contact is presented by the integral non-local relation between the electric induction **D** and the electric field **E**:

$$D_{m, \alpha}(\mathbf{r}) = \sum_{\beta} \int_{V_m} \varepsilon_{m, \alpha\beta}(\mathbf{r},\mathbf{r'}) E_{m, \beta}(\mathbf{r'}) d\mathbf{r'}, \qquad \alpha, \beta = x, y, z \qquad (2)$$

where m = 1, 2 refers to the media, solvent, and solute; the function $\varepsilon_{m,\alpha\beta}(\mathbf{r}, \mathbf{r'})$ is the dielectric permittivity tensor that is determined by the spatial correlation induced by the polarization of the medium; and the integration is taken over the volume $V_m$ of the medium. The main purpose of this approach is to incorporate the short-range structure of the contacting media into electrostatics.

We used the planar dielectric boundary to model the interface between dielectric (ceramic) including a protein-like medium (dielectric solute) and solvent (i.e. both media are semi-infinite). Integrations were performed using a cylindrical coordinate system, (R, Z), as shown in Figure 1, where Z is the axis perpendicular to the plane passing through the boundary, and R is the radius vector in this plane. The semi-infinite regions $Z < 0$ and $Z > 0$ were assigned to a solvent and a solute, respectively. The charge $Q = \xi e$ ($\xi$ is the fraction of the electron charge $e$) is located in the solute at point $(0, Z_0)$.

In the framework of the so-called "specular reflection approximation" model for the system of two semi-infinite media [32, 54] which we adopted here, the properties of the media along the plane of the boundary are considered as homogeneous isotropic with $\varepsilon_{m,\alpha\beta}(\mathbf{r}, \mathbf{r'}) = \varepsilon_{m,\alpha\beta}(Z, Z', \mathbf{R-R'})$ and can be expressed through the bulk dielectric function[41]. The differential form of Gauss's law[42] was applied with non-local relationship Eq.(2) in order to compute the electrostatic potential $\varphi(R, Z, Z_0)$ created by charge Q in any point $(R, Z)$ with $Z > 0$. The potential is expressed in terms of the Fourier-transformed dielectric functions $\varepsilon_1(\mathbf{r-r'})$ and $\varepsilon_2(\mathbf{r-r'})$ ($\varepsilon_1(k)$ and $\varepsilon_2(k)$, respectively), which characterize the bulk dielectric properties of the solvent and solute, respectively[9,32,54].

The corresponding image potential IP, $W(Z_0)$, is defined as a change in free energy of the system when transferring charge Q from infinity (bulk of the solute) to the point $R=0$, $Z=Z_0$ at the interface. It was calculated using Fourier-Bessel transformation in the Güntelberg charging cycle approach[43, 54]:

$$W(Z_0) = \frac{Q}{4\pi} \lim_{Z \to Z_0} \int_0^{+\infty} dKK \left[\varphi(K,Z,Z_0) - \varphi^0(K,Z,Z_0 \to \infty)\right], \qquad (3)$$

where $\varphi^0(K, Z, Z_0 \to \infty)$ denotes the electrostatic potential when the charge is located at infinity. In order to analyze the behavior of the IP potential for ceramics and protein, the dielectric solute was considered as a uniform dielectric medium with the dielectric constant $\varepsilon_2(k) = \varepsilon_C$. In the case of proteins, $\varepsilon_2(k) = \varepsilon_p = 4$[44]. For several typical ceramics, $\varepsilon_C$ is determined as a specific dielectric (low-frequency mode) constant: polyethylene, $\varepsilon_C = 2$; $SiO_2$, $\varepsilon_C = 3.9$[45]; quartz, $\varepsilon_C = 4.43$; $\varepsilon_C$ of $Ta_2O_5$ is in the range of 20-52[46,47,48]; $\varepsilon_C$ of zirconia in monoclinic phase is ~20-28[49], and similar values for its cubic phase stabilized by yttria[50]. The rutile phase of $TiO_2$ (also a phase of native surface oxide layer) has $\varepsilon_C \sim 114$[51,52,53]. The static dielectric permittivity of $TiO_2$ anatase phase was originally reported to be ~31[51]. Fukushima and Yamada have shown that the dielectric constant of anatase $TiO_2$ may vary between 35 and 108 depending on preparation conditions[18].

The dielectric function $\varepsilon_1(k)$ in the bulk of the aqueous solvent was approximated in the context of polar solvent phenomenological model [32, 40, 54]:

$$\varepsilon_1(k) = \varepsilon_* + (\varepsilon_s - \varepsilon_*)/[1 + (Lk)^2 \varepsilon_s / \varepsilon_*], \qquad (4)$$

where $\varepsilon_* = 6$ and $\varepsilon_s = 78.3$ are short- and long-wavelength dielectric constants of the solvent at room temperature; and L is the correlation length of the water dipoles, which is proportional to the characteristic length of the hydrogen-bonding network of water molecules (~3-5Å). The dielectric function $\varepsilon_1(k)$ is the "one pole approximation" which effectively separates the spectrum of the polarization fluctuations of water into two parts: (i) the low-frequency zone associated with the orientational Debye mode; and (ii) the high-frequency zone associated with infrared and optical modes. This approximation takes into account the orientational Debye polarization and neglects the higher frequency modes[32, 55,56,57]. The polarization modes within the orientational mode are correlated in space with effective correlation radius L (due to the strong non-electrostatic interactions), while the polarization modes with the high-frequency are assumed to be uncorrelated [55,58]. The dielectric constant, determined within the transparency zone of the electromagnetic absorption (which separates frequencies typical for low- and high-frequency zones), corresponds to the short-wavelength dielectric constant $\varepsilon_*$ [55,58]. Thus, the "one pole approximation" dielectric model for water assumes that the function $\varepsilon_l(k)$ is changed at the length-scale $\sim L$ from the values characteristic for the macroscopic, long-wavelength dielectric constant, $\varepsilon_s$, to the value of the short-wavelength dielectric constant $\varepsilon_* \cong 6$ [55,56,58]. The "one pole approximation" was found to be adequate for the description of polar molecules without internal degrees of freedom, such as rigid, strongly correlated dipoles[55].

Quantities $\varepsilon_*$ and L of water can be adjusted/selected by fitting to the experimental data analyzed in terms of the free energy of interaction between protonated amino groups in dibasic amines [57]. Function $\varepsilon_1(k)$ in Eq.(4) was applied in the framework of the NL electrostatic approach to explain several experimental data in the electrolyte theory, interfacial electrochemistry, and computational biophysics[9,32,40] (and references cited therein).

The electrostatic free energy of the dielectric solute (ceramic and protein) in the unbound states $\Delta G^C$ and $\Delta G^P$ was calculated using Eq.(3). $W(Z_0)$ potential in the $k_BT$ energy units can be written in the form of[9]

$$W(Z_0)/(560 \, k_BT \, \xi^2) = (4\varepsilon_d Z_0)^{-1} \int_0^{+\infty} dx \, S(x, \eta) \, \exp(-x), \quad (5)$$

$$S(x, \eta) = [D(x, \eta) - 1/\varepsilon_d] / [D(x, \eta) + 1/\varepsilon_d],$$

$$D(x, \eta) = (1/\varepsilon_S) + (1/\varepsilon_* - 1/\varepsilon_S)[1 + (x\eta)^{-2}]^{-1/2},$$

where $\eta = L/2Z_0$ is a dimensionless parameter, $\varepsilon_d$ is solute dielectric constant and $Q^2 = \xi^2 e^2 = 560 \, k_BT \, \xi^2$.

The results of the asymptotic and numerical analysis of the above IP potential [9] and pair wise electrostatic interaction energy at the interface [32,40] suggest that a layer of interfacial water on the solute dielectric surface has effective (i.e. distance- dependant) dielectric properties different from the bulk solvent. The corresponding effective dielectric permittivity of the above interfacial solvent layer was calculated as a function of distance from the interface[9,40]. The magnitude of the function is $\sim (\varepsilon_d + \varepsilon_*)/2$ at small $Z \sim Z_0 < L$ (in close proximity to the interface), while it approaches $\sim (\varepsilon_d + \varepsilon_S)/2$ at the large distances $Z \sim Z_0 > 10\text{-}15$ Å. The validity of these average dielectric constants at the interface is also supported by recent work[59]. It should be noted that the occurrence of a low-dielectric layer on the dielectric surface is consistent with the experimental study of hydration dynamics on the protein surface [34,40] (and references cited therein).

**RESULTS AND DISCUSSION**

The electrostatic binding energy between protein and ceramic (two solutes), forming the complex, was calculated using the above model of idealized (planar) interface of proteins P and ceramic C. The protein possesses a single ion charge (for example $+e$) located at the protein surface interacting patches or protein "binding site". Analogously, the ceramic also possesses a single electron charge of opposite sign ($-e$) located at the ceramic surface interacting patch ($\xi_P = \xi_C = 1$). Both charges were considered as a point charge located in the solute (P and C media) in the proximity of interface. The bound state of the CP complex occurs through the contact of the corresponding binding sites (facing each other) of protein P and ceramic C. The point charges of the binding sites in the bound pair are assumed to have a minimal inter-charge distance of $r_{12}$. This distance is about $r_{12}= 2r_{ion} \approx 5$ Å, where $r_{ion} \approx 2.5$ Å is the closest proximity of a charged center of an ionogenic group of amino acid residue to the molecular protein surface. The size of the adhesive protein is usually much larger than ionic radius, thus, justifying the planar interface approximation in the vicinity of the charge to estimate the electrostatic component of the free energy of binding in the bound state. However, when the size of protein is comparable to the ionic radius, the curvature of the interface cannot be neglected and corresponding correction needs to be included[60,61]. The similar planar approximation can be used for the smooth ceramic surfaces when the local curvatures of its surface features are small compared with inverse inter-atomic distances in the ceramic (~2-3Å).

The electrostatic free energy of the bound system in planar interface approximation is described by the first term of Eq.(1). This is a Coulomb interaction energy at the interface of the CP complex between the protein and ceramic with dielectric constant $\varepsilon_P$ and $\varepsilon_C$, respectively, and the inter-charge distance of $r_{12}$ mentioned above [42]:

$$\Delta G^{CP}/(560\ \xi_P\xi_C k_B T) = -1/(\varepsilon_{eff}\ r_{12}), \qquad (6)$$
$$\varepsilon_{eff}= (\varepsilon_P + \varepsilon_C)/2,$$

where $k_B T \approx 0.593$ kcal/mol at 25°C.

We calculated the electrostatic energy of the unbound states $\Delta G^C$ and $\Delta G^P$ as the IP potential, Eq.(3). All calculations were performed with a typical distance $Z_0=2.5$Å. The behavior of the $\Delta G^C$ for the charge located in the dielectric media (ceramic) as a function of the media dielectric constant $\varepsilon_C$ considered in NL ($\varepsilon_* = 6$, $\varepsilon_s = 78.3$, L = 5 Å) and CL ($\varepsilon_S = \varepsilon_* = 78.3$) solvent approximation is shown in Figure 2. As we can see in Figure 2, the results of numerical calculations obtained for Eq.(5) are consistent with result of our asymptotic analysis carried out in the CL case [42]:

$$\frac{\Delta G^C_{CL}(\varepsilon_C, Z_0)}{560\xi^2 k_B T} = \frac{1}{4Z_0\varepsilon_C}\frac{\varepsilon_C - \varepsilon_S}{\varepsilon_C + \varepsilon_S}, \quad Z_0 > 0,$$

and NL consideration[9]

$$\frac{\Delta G^C_{NL}(\varepsilon_C, Z_0)}{560\xi^2 k_B T} = \frac{1}{4Z_0\varepsilon_C}\frac{\varepsilon_C - \varepsilon_*}{\varepsilon_C + \varepsilon_*}, \quad Z_0 \ll L.$$

The asymptotic and numerical analyses of the $\Delta G^C$ indicate this energy is negative (attractive interactions) at the small values of the ceramic dielectric constant ($\varepsilon_C < \varepsilon_*, \varepsilon_S$), while $\Delta G^C$ has a positive value (repulsive interactions) at the large values ($\varepsilon_C > \varepsilon^*, \varepsilon_S$) both in CL and NL consideration (Figure 2). The previously found significant decrease of dielectric permittivity of the solvent at the interface is responsible for this effect[9]. The corresponding value of $\Delta G^P$ for the charge located in the protein ($\varepsilon_P=4$, $Z_0=2.5$Å) was calculated as $\Delta G^P=-12.6 k_B T$ for CL and $\Delta G^P = -6.6 k_B T$ for NL solvent consideration[9].

The electrostatic free energies $\Delta G_{NL,EI}$ and $\Delta G_{CL,EI}$ of the model protein-ceramic binding accordingly for NL and CL solvent consideration were calculated as a function of the ceramic dielectric constant $\varepsilon_C$ using Eq.(1), where $\Delta G^{CP}$ is calculated by Eq.(6), and $\Delta G^C$ and $\Delta G^P$ - by Eq.(5) with $\varepsilon_d=\varepsilon_C$ and $\varepsilon_d=\varepsilon_p=4$, respectively. The numerical calculation of the $\Delta G_{EI}$, $\Delta G^{CP}$, and $\Delta G^C$ energies for several typical dielectrics are summarized in Tables 1 and 2 for the NL and CL approximation, respectively. Figure 3 shows the electrostatic contribution to the binding free energy of protein-ceramic association ($\Delta G_{NL,EI}$) is shifted significantly (to the negative values) when we apply NL electrostatic solvent model compared to the CL one. As it follows from obtained results (Figure 3), the electrostatic free energy of the protein–surface complex calculated using CL consideration of the aqueous solvent is unfavorable for most of metal oxide ceramics. At the same time, consideration of the solvent by the NL electrostatic model reveals significant strengthening of the electrostatic contribution to the binding energy making the protein-surface complex formations electrostatically favorable for dielectric constant $\varepsilon_C<35$. In this case, the complex formation remains unfavorable for $\varepsilon_C$ over 35, but $\Delta G_{EI}$ is considerably reduced. Origin of the above effect reflects the significant increase of $\Delta G^C$ and $\Delta G^P$ energies due to presence of a low-dielectric solvent layer on the ceramic and protein surface in the process of the protein-ceramic association in the solvent (see Tables 1 and 2). It should emphasize that $\Delta G^P$ is a negative value and provides the same significant shift for $\Delta G_{EI}$ at all $\varepsilon_C$ values. The energy $\Delta G_{EL}$ is a non-monotonic function of $\varepsilon_C$, and $\Delta G_{EL}$ increases smoothly as a function of $\varepsilon_C$ in both CL and NL solvent consideration from minimum at $\varepsilon_C=4$. This is because the pairewise Coulomb interaction energy at the interface $\Delta G^{CP}$, Eq.(6), is negative and inversely proportional to $(\varepsilon_P +\varepsilon_C)$, while $\Delta G^C$, Eq.(5), is positive at large $\varepsilon_C$ (as it is following from Figure 2 and asymptotic analysis). At the small $\varepsilon_C$, $\Delta G^C$ becomes negative (Figure 2) and increases rapidly in magnitude when $\varepsilon_C$ decreases. As a result, $\Delta G_{EI}$ becomes positive (repulsive) at very small $\varepsilon_C$ (Figure 3).

Table 1. The numerical calculation of the $\Delta G^C$, $\Delta G^{CP}$ and $\Delta G_{EI}$ energies for several typical dielectrics in the case of the non-local (NL) electrostatic model of the solvent

| Dielectric materials | Dielectric constant, $\varepsilon_C$ | $\Delta G^C$ [$k_B T$] | $\Delta G^{CP}$ [$k_B T$] | $\Delta G_{NL,EI}$ [$k_B T$] |
|---|---|---|---|---|
| Polyethylene | 2 | -19.29 | -37.33 | -11.42 |
| Quartz | 4 | -6.62 | -28.00 | -14.76 |
| ZrO2 | 25 | 0.71 | -7.72 | -1.81 |
| Ta2O5 | 30 | 0.73 | -6.59 | -0.70 |
| TiO2 | 110 | 0.39 | -1.96 | +4.27 |

Table 2. The numerical calculation of the $\Delta G^C$, $\Delta G^{CP}$ and $\Delta G_{EI}$ energies for several typical dielectrics in the case of the classical (CL) electrostatic model of the solvent

| Dielectric materials | Dielectric constant, $\varepsilon_C$ | $\Delta G^C$ [$k_B T$] | $\Delta G^{CP}$ [$k_B T$] | $\Delta G_{CL,EI}$ [$k_B T$] |
|---|---|---|---|---|
| Polyethylene | 2 | -26.60 | -37.33 | 1.9 |
| Quartz | 4 | -12.63 | -28.00 | -2.74 |
| ZrO2 | 25 | -1.16 | -7.72 | +6.07 |
| Ta2O5 | 30 | -0.83 | -6.59 | +6.87 |
| TiO2 | 110 | 0.085 | -1.96 | +10.59 |

The ceramics with dielectric constants below a specific value (in our estimations ~35) favor the association of protein and surface into a complex due to the binding energy $\Delta G_{EI}$ (calculated by the NL solvent model) becomes negative in comparison with CL consideration (Figure 3). Thus, $\Delta G_{EI}$ calculated for a protein complex with the $ZrO_2$ surface is more favorable than for ceramic surfaces with larger dielectric constants (e.g. $TiO_2$: Tables 1). In fact, the low-dielectric boundary water layer on the protein and ceramic surface is responsible for favorable protein-ceramic complex formation ($\Delta G_{EI}<0$) for several ceramics indicated in Figure 3 by their respective typical low-frequency dielectric constants. Overall, our estimations show that the interfacial solvent layer with low dielectric permittivity on the both protein and ceramic surfaces is a major factor capable of compensation for the unfavorable desolvation effects in the protein adsorption on the dielectric implant surface.

It should be noted that the above consideration takes into account only a single oppositely-charged pair across the implant-protein interface. For this pair, for example in case of zirconia, ($\Delta G_{EI}$) is ~ -2$k_B$T (Table 1). At the same time, there are many oppositely-charged pairs at the protein-ceramic interface resulting in strong electrostatic attractive interaction. Thus, complimentary EI can be a driving force for protein adsorption and formation of protein-implant complexes. The formation of multiple charge patches on the ceramic surface can be achieved by surface modification such as patterning and/or nano-structuring. This is supported by studies showing that a sharper tantalum oxide curvature nanostructure promotes the adsorption and activation of adhesive protein such as fibronectin.[62]

The electrostatic binding between ceramic and adhesive protein can be strengthened by reducing the dielectric constant of the ceramic, and at the same time, increasing the concentration of surface-charged nanostructures. Based on our model calculations, we deduced that $ZrO_2$ or $Ta_2O_5$ coatings could have advantages with respect to $TiO_2$ or a native oxide on the surface of Ti, and provide stronger electrostatic binding. Furthermore, the co-alloying of these compounds with $TiO_2$ and reducing dielectric permittivity of the coating[18] can also strengthen the electrostatic component to the binding energy. Furthermore, nanostructuring of the $TiO_2$ coating promotes the formation of the anatase phase with a reduced dielectric constant ($\varepsilon_C$~30) may serve the same purpose. These results are in qualitative agreement with the studies of the fibronectin adsorption on various ceramic surfaces [62,63,64].

**Conclusion**

In the present work, we calculated and analyzed the contribution of electrostatic interactions (EI) to the free energy of binding between model protein and ceramic implant surface in the aqueous solvent as a function of the ceramic (solid implant) dielectric constant. We used non-local electrostatic approach[9, 32, 40] to take into account the solvent structure and the contribution of a solvent orientational polarization (correlated by the network of hydrogen bonds) into the pairwise EI at the considered interfaces with solvent.

We show that the existence of the low-dielectric boundary water layer at the protein-solvent ("dynamically ordered water") [32, 34] interface and ceramic-solvent interface markedly strengthens the electrostatic contribution to the protein-ceramic binding energy ($\Delta G_{EI}$) due to the increase of charging energy of the protein and ceramic implant ($\Delta G^C$ and $\Delta G^P$ energies) in aqueous solvent. Using the non-local electrostatic approach, we show the protein-implant association is favorable (attractive) for implant dielectric constant $\varepsilon_C<35$, while consideration of the aqueous solvent as a uniform dielectric medium with high dielectric constant $\varepsilon_s = 78.3$ (that is typically assumed in classical local approach) predicts unfavorable (repulsive) EI between protein and a majority of typical metal oxide ceramic materials ($\varepsilon_C>10$).

The solid implants coated by oxide ceramics with reduced dielectric constants can be beneficial to strengthen the electrostatic binding of the adhesive proteins to the implant. In particular, we found that electrostatic binding energy for protein adsorption on the $ZrO_2$ or $Ta_2O_5$ surface is more favorable than for ceramic surfaces with a larger dielectric constant including known biocompatible materials such as $TiO_2$. Alternatively, metals such as Ta and Zr can improve implant biocompatibility, because their native oxide has lower dielectric constant. Such control of the protein adsorption at the surface is paving the way for development of a novel coating for orthopaedic implants.

**Acknowledgement**

The work at the University of Nebraska Medical Center in Omaha, NE, was supported by DOE grant, "Material Science Smart Coatings", Award No. DE-SC0005318. The work at the University of Nebraska at Omaha was supported by NSF-EPSCoR (Grant No. EPS-1010674), and DOE DE-EE0003174, and Nebraska Research Initiative.

**Figure captions.**

Figure 1. The charge Q located at the point (0, $Z_0$) of the cylindrical coordinate system, $Z_0$ is the distance to the interface.

Figure 2. The numerical calculation of the electrostatic energy of the unbound states $\Delta G^C$ for the charge located in the dielectric media (ceramic) as a function of the media dielectric constant for NL (Red) and CL (Blue) solvent consideration. Parameters of the solvent for the curves: (Blue) $\varepsilon_S = \varepsilon_* = 78.3$; (Red) $\varepsilon_* = 6$, $\varepsilon_s = 78.3$, L = 5 Å. The values of the dielectric constants of the ceramic materials are: polyethylene: $\varepsilon_C= 2$; quartz: $\varepsilon_C= 4$; $ZrO_2$: $\varepsilon_C= 25$; $Ta_2O_5$: $\varepsilon_C=\sim 30$; $TiO_2$: $\varepsilon_C = 114$ (rutile). Calculations were performed with $Z_0=2.5$Å.

Figure 3. The electrostatic free energy $\Delta G_{EL}$ of the model protein-ceramic binding for non-local $\Delta G_{NL,EI}$ (Red) and classical $\Delta G_{CL,EI}$ (Blue) solvent consideration in depending on ceramic material dielectric constant $\varepsilon_C$. Calculations were performed with $Z_0=2.5$Å.

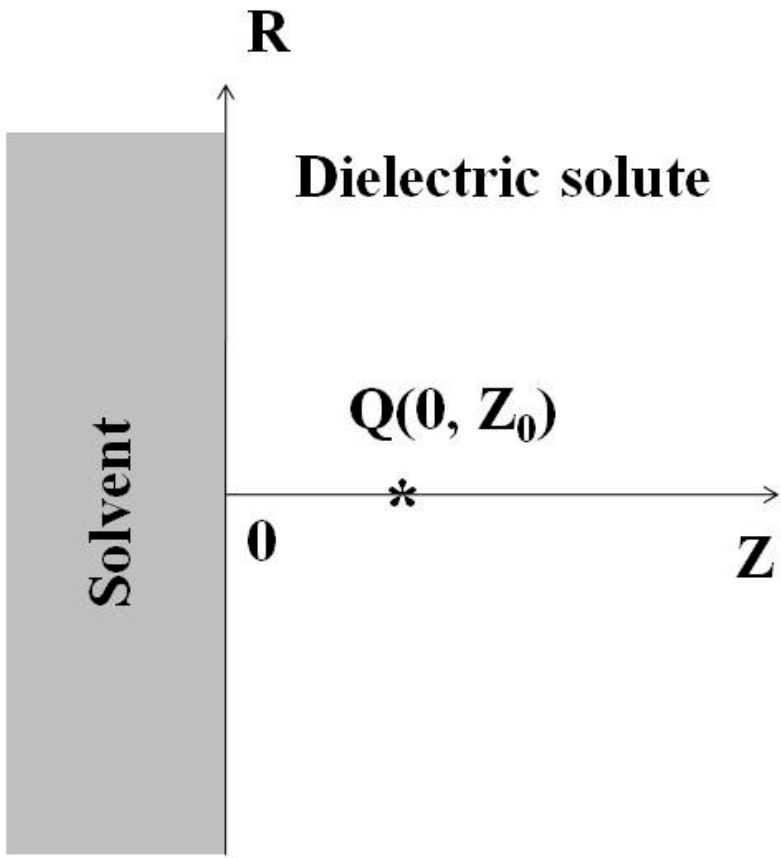

Figure 1

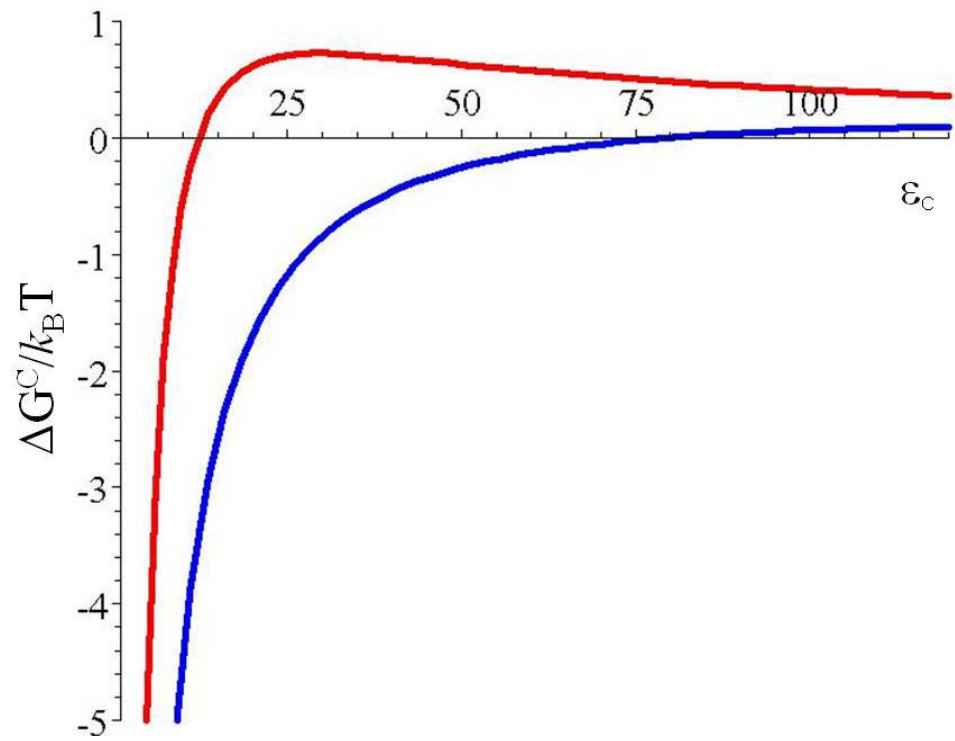

Figure 2.

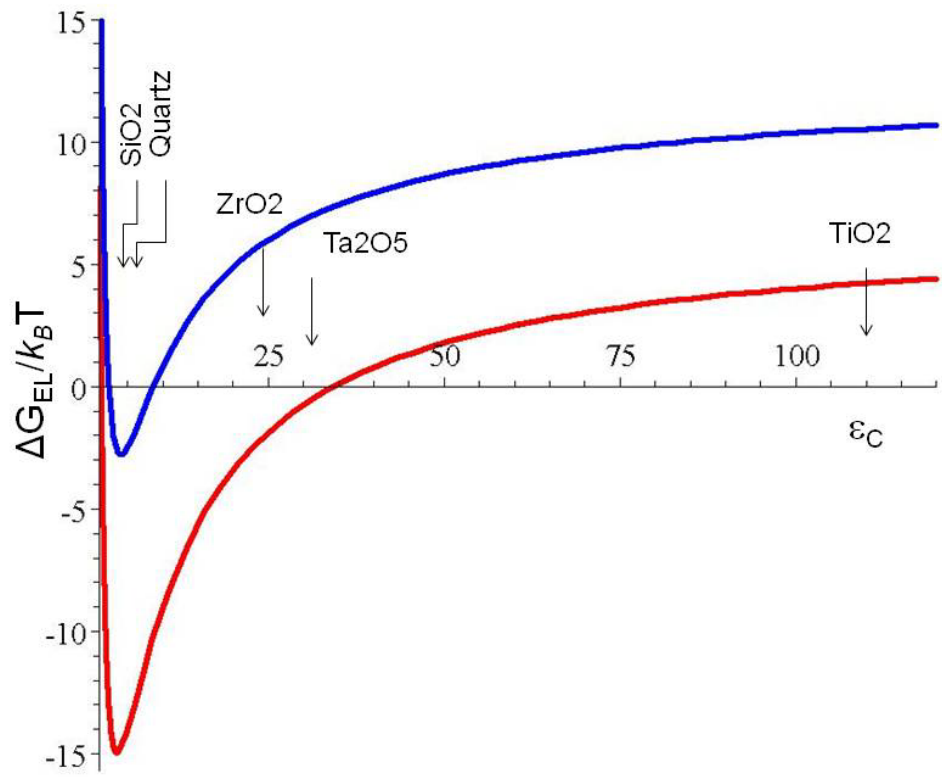

Figure 3